\documentclass[final,3p]{elsarticle}

 \usepackage{graphics}
 \usepackage{graphicx}
 \usepackage{booktabs}
 \usepackage{makecell}
\usepackage{multirow}
\usepackage{caption}
\usepackage{subcaption}
\usepackage{amssymb,amsthm}
\usepackage{amsmath}
\usepackage{color}
\usepackage{soul}           
\usepackage[english]{babel}
\usepackage[nottoc]{tocbibind}
\usepackage{ulem}
\usepackage{tikz}
\usepackage{bm}
\usetikzlibrary{shapes.geometric, arrows}

\tikzstyle{normal} = [rectangle, rounded corners, minimum width=3cm, minimum height=1cm,text centered, draw=black, fill=red!30,text width=10em]

\tikzstyle{robust} = [trapezium, trapezium left angle=70, trapezium right angle=110, minimum width=3cm, minimum height=1cm, text centered, draw=black, fill=blue!30,text width=4em]
\usepackage{amssymb}
\usepackage{amsmath}
\usepackage{color}

\newtheorem{definition}{Definition}
\newtheorem{theorem}{Theorem}
\newtheorem*{proof*}{Proof}

\newtheorem{example}{Example}

\usepackage[shortlabels]{enumitem}
\usepackage{algorithm,algpseudocode}
\usepackage{hyperref}
\hypersetup{
colorlinks=true
}
\usepackage{epstopdf}

\usepackage[T1]{fontenc}
\usepackage{hyperref}
\hypersetup{
colorlinks=true
}
\usepackage{epstopdf}
\usepackage{xcolor}

\biboptions{sort&compress}

\journal{arXiv}

\begin{document}

\begin{frontmatter}

\title{{Identification and validation of  periodic autoregressive model with additive noise: finite-variance case}}

\author[label1]{Wojciech Żuławiński}\corref{cor1}\ead{wojciech.zulawinski@pwr.edu.pl}
\cortext[cor1]{Corresponding author.}
\author[label1]{Aleksandra Grzesiek}
\author[label2]{Rados{\l}aw Zimroz}
\author[label1]{Agnieszka Wy\l oma\'nska}

\address[label1]{Faculty of Pure and Applied Mathematics, Hugo Steinhaus Center, Wroclaw University of Science and Technology, Wyspianskiego 27, 50-370 Wroclaw, Poland}
\address[label2]{Faculty of Geoengineering, Mining and Geology, Wroclaw University of Science and Technology, Na Grobli 15, 50-421 Wroclaw, Poland }

\begin{abstract}
In this paper, we address the problem of modeling data with periodic autoregressive (PAR) time series and additive noise. In most cases, the data are processed assuming a noise-free model (i.e., without additive noise), which is not a realistic assumption in real life. The first two steps in PAR model identification are order selection and period estimation, so the main focus is on these issues. Finally, the model should be validated, so a procedure for analyzing the residuals, which are considered here as multidimensional vectors, is proposed. Both order and period selection, as well as model validation, are addressed by using the characteristic function (CF) of the residual series. The CF is used to obtain the probability density function, which is utilized in the information criterion and {for residuals distribution testing.} To complete the PAR model analysis, the procedure for estimating the coefficients is necessary. However, this issue is only mentioned here as it is a separate task (under consideration in parallel). The presented methodology can be considered as the general framework for analyzing data with periodically non-stationary characteristics disturbed by finite-variance external noise.  The original contribution is in {the selection} of the optimal model order and period identification, as well as the analysis of residuals. All these findings have been inspired by our previous work on machine condition monitoring that used PAR modeling
\end{abstract}


\end{frontmatter}
\section{Introduction}

{In this paper, we propose methods for identifying periodic autoregressive (PAR) models in data that consist of pure PAR time series and additive noise. Our focus is on the case of finite-variance, meaning that both the innovation series and the additive noise {have finite variance}. The PAR time series is considered one of the most common models with periodically correlated (PC) property, which is often observed in systems that have a periodically non-stationary random structure \cite{napolitano2016cyclostationarity,hurd2007periodically}. It has found many interesting applications; see, e.g., hydrology \cite{hyd2}, climatology and meteorology \cite{met1}, economics \cite{broszkiewicz2004detecting}, and condition monitoring \cite{Antoni2002815,Wodecki201989,Sun2020,Wodecki2021}. In the classical literature, mostly the pure (noise-free) PAR model is discussed. However, this situation seldom occurs in real-world applications, and the measured signal is almost always corrupted by noise.} {In practice, the additive noise can be Gaussian, contain outliers, or even be leptokurtic. This requires specific procedures that differ from those dedicated to a pure model.}
{The research studies on autoregressive  (AR) model identification with additive noise with different types of additional disturbances are {well-known} in the literature  \cite{esfandiari,diversi1,MAHMOUDI2010157,CAYIR2021108118,MAHMOUDI20121151, MAHMOUDI20111659,HASAN2003603,MAHMOUDI20082777,DIVERSI20072843,LABARRE20062863}}.  
In recent research, one can also find studies on the PAR model with additive noise.  However, most of the attention is focused on the case {where} the additive noise consists of a sequence of additive outliers, which are independent and identically distributed (i.i.d.) random variables with large values that occur with a certain probability. We recall here the bibliography references where new estimation techniques are proposed for such models \cite{parma_ao0,parma_an1,parma_ao2,parma_ao3,parma_ao4}. However, these algorithms do not account for the presence of additive disturbances and instead use robust estimators of classical statistics to minimize the effect of large observations that can be seen in the signal. The special case, namely the PAR(1) model with general finite-variance additive noise, is discussed in \cite{nasza_wojtek}. 
Therefore, it can be concluded that the identification of the PAR model with general additive noise is rarely addressed in the literature and there is a space for developing new and effective approaches. This aspect has significant practical importance

{The procedure of identifying a PAR model can be structured into several stages: selecting the model order, determining the period, estimating the model coefficients, and conducting a {residual} analysis to validate the model. In certain situations, the model order and period may be advised or provided by other sources. The most challenging aspect is finding the values of {the} coefficients to complete the model, which is a serious problem from both a theoretical and algorithmic perspective and has been discussed separately. In this paper, we briefly recall the modified errors-in-variables method (see \ref{appB}). 
The final step is to validate the model by analyzing its residuals, which could also be considered as identifying an unknown distribution.} 

As mentioned {above}, the initial step in analyzing real data is selecting the optimal order of the model and determining its period (relevant for PC models). For pure models, the model order identification problem can be solved using commonly known information criteria, such as $BIC$ and $AIC$ (see \cite{aicc,aicc1}), which are calculated under the assumption of a specific distribution of the residual series. However, when additive noise is present, these classical approaches may not be sufficient. In this paper, the proposed methods for optimal model order and period identification are also based on information criteria. However, unlike classical approaches, the residuals of the analyzed noise-corrupted PAR model are dependent. To overcome this issue, we analyze the residuals as $T$-dimensional vectors, where $T$ is the period of the PAR model. This arrangement in vectors allows us to consider the residuals as i.i.d $T$-dimensional random variables with dependent components. Then, when we assume that both the innovation series of the PAR model and the additive noise are Gaussian, it can be deduced that the residuals also have a $T$-dimensional Gaussian distribution with appropriate parameters. However, the situation becomes much more complicated when the additive noise is not Gaussian, as is the case in some real-world applications.

 In such a case, we propose {to apply} the inverse of the characteristic function (CF) for {the} residuals to obtain their probability density function (PDF), which will be used in the information criterion. The theoretical CF is provided in this paper for $T$-dimensional vectors representing the residuals under the general assumption of a finite-variance PAR model with additive noise. The form of the CF is also utilized in the final step of the data analysis, which is the validation of the model. We propose an approach to test the distribution of the residuals by comparing the empirical CF of the residual series with the theoretical CF of the tested distribution. A similar methodology has been applied, for instance, in \cite{Sabuka} for sub-Gaussian distributions.

The proposed algorithms were confirmed to be efficient through extensive Monte Carlo simulations for two cases of additive noise distribution: Gaussian and a mixture of Gaussian distributions (which represents the leptokurtic class of distributions). The optimal order identification was considered in two scenarios, with known and unknown period $T$. In both cases, the performed simulations clearly indicate the efficiency of the proposed methodology. We note that the presented methodology for order and period identification, estimation of the parameters, and validation of the model by {residual} analysis, can be considered as {a} general framework for analysis of the data with periodically non-stationary characteristics disturbed by general finite-variance external noise. We believe that this methodology will find {applications} in many areas of interest.

The rest of the paper is organized as follows. In Section \ref{sec:model}, we introduce the {noise-corrupted}  finite-variance PAR model. In Section \ref{ident}, we describe the procedure for the identification of  the optimal order and period for the considered model. Next, in Section \ref{sec:val}, we introduce {a} procedure for {the residual} analysis to validate the fitted model. In Section \ref{sec:simul}, we present {a} simulation study  for the optimal $p$ and $T$ {value} selection and for {testing the distribution of residuals}. The last section concludes the paper.
{{Note that in \ref{appaa} we demonstrate the equivalent matrix representation of the residual series.} In \ref{appa}, we recall the class of mixture of Gaussian {distributions}, {while in \ref{appbb} we present the alternative proof for residuals distribution in the case where the additive noise is a mixture of Gaussian distributions. Finally,} in \ref{appB}, we {provide more details connected with the third step} of the mentioned procedure, i.e., {estimating the model coefficients}{.} {However, a more detailed study regarding this subject is} prepared as a separate material.}
\section{Finite-variance periodic autoregressive model with additive noise} \label{sec:model}

The periodic autoregressive model of order $p$ with additive noise (also called noise-corrupted PAR($p$) model) is defined as follows:
\begin{eqnarray}\label{model1}
Y_t=X_t+Z_t,
\end{eqnarray}
where $\{X_t\}_{t\in \mathbb{Z}}$ is {a} periodic autoregressive (PAR) time series of order $p\in \mathbb{N}$ and period $T\in \mathbb{N}$ defined below. {For simplicity}, we assume {that} $p<T$. {However, the case $p\geq T$ is briefly treated as well}. We assume that  the innovations  of the PAR($p$) model (denoted further  as $\{\xi_t\}_{t\in\mathbb{Z}}$) and the additive disturbances $\{Z_t\}_{t\in\mathbb{Z}}$ are independent {sequences of i.i.d. random variables with mean zero} and variances $\sigma_{\xi}^2$ and $\sigma_{Z}^2$, respectively. 
In this paper, we consider the additive noise as a sample of i.i.d. random variables from a continuous distribution with PDF $f_Z(\cdot)$. However, the presented methodology is universal and can also be applied in the case where the sequence $\{Z_t\}$ is derived from {a} discrete distribution. In this case, the additive disturbances are considered as  additive outliers \cite{parma_ao0}. 

\begin{definition}\cite{vecchia1985periodic}
\label{def:parma1}
{A} sequence $\{X_t\}$ is a finite-variance PAR($p$) model with period $T\in \mathbb{N}$ {if} it satisfies the following equation{:}
\begin{eqnarray}\label{parma1}
X_t-\phi_1(t)X_{t-1}-\cdots-\phi_p(t)X_{t-p}=\xi_{t}.
\end{eqnarray}\end{definition}
In Eq. (\ref{parma1}){,} it is assumed that the parameters $\phi_1(t),\phi_2(t),\cdots,\phi_p(t)$ are periodic in $t$ with the same period $T$. It is common to consider  $\{\xi_t\}$ {to be} a sequence of  Gaussian distributed random variables. In simulation studies, we also consider this case. However, the presented methodology is also valid for any finite-variance distribution of the innovation series. We mention that, {similarly to} the PAR($p$) time series, the noise-corrupted model $\{Y_t\}$ exhibits the periodically correlated (PC) property, which means that its mean and autocovariance functions are periodic with respect to $t$ with the same period $T$ \cite{hurd2007periodically}. More precisely, we can show that
\begin{eqnarray}
\mathbb{E}Y_t=\mathbb{E}Y_{t+T},~~cov(Y_t,Y_{t+h})=cov(Y_{t+T},Y_{t+h+T}),~~h\in \mathbb{Z}.
\end{eqnarray}
For {a} detailed proof of the special case, namely for the model with $p=1$, we refer the readers to \cite{nasza_wojtek}. 

One can easily show that for each $t\in \mathbb{Z}$ the time sequence $\{Y_t\}$ satisfies the following equation: 
\begin{eqnarray}\label{y1}
Y_t-\phi_1(t)Y_{t-1}-\cdots -\phi_p(t)Y_{t-p}=\xi_t+Z_t-\phi_1(t)Z_{t-1}-\cdots-\phi_p(t)Z_{t-p}.
\end{eqnarray}
The residuals $\{R_t\}$ of the model given in Eq. (\ref{model1}) are as follows:
\begin{eqnarray}\label{res00}
R_t=Y_t-\phi_1(t)Y_{t-1}-\cdots -\phi_p(t)Y_{t-p}.
\end{eqnarray}
Obviously, in the considered case, the residuals are not independent. However, by defining the $T$-dimensional time series
\begin{eqnarray}\label{res11}
\mathbf{R}_n=[R_{nT+1},\cdots,R_{(n+1)T}],~~ n=1,2,\cdots,N-1, 
\end{eqnarray}
one obtains the vectors of $T$-dimensional i.i.d. random variables (with dependent components). {In  \ref{appaa}, the  matrix representation of the residual series is presented. This representation is equivalent to the one given in this section and can be useful e.g. for the identification of the residuals distribution. }
One can show that $
\mathbb{E}(\mathbf{R}_n\mathbf{R}_m')=0$, $n\neq m$ ($A'$ denotes the transposition of {the} matrix $A$). Moreover, the covariance matrix ${\mathbf{\Gamma}^R}=[{\gamma^R}(k,l)]_{k,l=1}^T$ of $\mathbf{R}_n$ is independent on $n$ and the coefficients ${\gamma^R}(k,l)=\mathbb{E}(R_{nT+k}R_{nT+l})$ are given by 
\begin{align} \label{gamma}
 {\gamma}^R(k,l) = \left\{ \begin{array}{ll}
  \sigma_{\xi}^2+\sigma_{Z}^2\sum_{j=0}^p\phi^2_j(k) & \textrm{if $k=l$},\\
   \sigma_{Z}^2\sum_{j=0}^{p+k-l}\phi_j(k)\phi_{j+l-k}(l) & \textrm{if $k<l$},\\
    \sigma_{Z}^2\sum_{j=0}^{p+l-k}\phi_j(l)\phi_{j+k-l}(k) & \textrm{if $k>l$},
    \end{array} \right. 
\end{align}
assuming that $\phi_{0}(k)=-1$  for any $k=1,2,\cdots,T$.
In the case where the sequences $\{\xi_t\}$ and $\{Z_t\}$  are Gaussian distributed, the PDF of $\mathbf{R}_n$ for any $n$ is given by
\begin{eqnarray}\label{pdf_R}
f_{\mathbf{R_n}}(\mathbf{r})=\frac{\exp\left\{-\frac{1}{2}\mathbf{r}(\mathbf{\Gamma}^R)^{-1}\mathbf{r}'\right\}}{\sqrt{(2\pi)^T|{\mathbf{\Gamma}^R|}}},
\end{eqnarray}
where $\mathbf{r}=[r_1,\cdots,r_T]$. {The fact that in this case the $\mathbf{R}_n$ is multivariate Gaussian with the PDF given in Eq. \eqref{pdf_R} can also be derived from a matrix representation presented in \ref{appaa}}. In the case {where} $p\geq T$, one can consider the same methodology as above, but for $KT$-dimensional residual blocks $\mathbf{R}_n$, where $K$ is the least natural number for which $p<KT$. In Section \ref{sec:val} we also present how to find the distribution of the residual series of the model (\ref{model1}) {if the sequence $\{Z_t\}$} is not Gaussian distributed. Let us note that the distribution of the $T$-dimensional sequence $\{\mathbf{R}_n\}$ is crucial for the validation of the model.

\section{Identification of the optimal $p$ and $T$ parameters}\label{ident}
For identifying the optimal order $p$ and period $T$ of the PAR($p$) model with additive noise given in Eq. (\ref{model1}), we propose to use the Bayesian information criterion ($BIC$), which is the estimator of the prediction error and, thereby, the relative quality of statistical models for a given set of data. However, {other well-known criteria} (such as $AIC$) could also be used in {a} similar manner.   Given a collection of fitted models for real data, $BIC$ estimates the quality of each model relative to other models. First, we present the case where the period $T$ of the PAR model is known. The idea of selecting the optimal order of the model $p^{*}_{opt}$ in this case is as follows. {For each order} $p^*=1,2,\cdots,p_{max}$ ($p_{max}<T$),  we estimate the parameters of the model given in Eq. (\ref{model1}). Here, we propose to use the estimation algorithm presented in \ref{appB}. Then we calculate the $BIC$ statistic{, defined as}
\begin{eqnarray}\label{bic_crit}
BIC=-2log(\mathcal{L})+log(\mathcal{N})\mathcal{K},
\end{eqnarray}
where  $\mathcal{L}$ is the likelihood function, $\mathcal{K}$ is the number of the estimated parameters, and $\mathcal{N}$ is the number of observations {(see \cite{aicc,aicc1})}. {Finally, the optimal order is equal to the $p_{opt}^{*}$  that minimizes the $BIC$ statistic in Eq. \eqref{bic_crit}}.

For the considered case, the residuals of the model, see Eq. (\ref{res00}), do not constitute a sample of independent random variables. Thus, the $BIC$ statistic is defined here for the series of $T$-dimensional i.i.d. random variables defined in Eq. (\ref{res11}). If we assume that for each $n$, $\mathbf{R}_n$ has PDF $f_{\mathbf{R}_n}(\cdot)$, then the $BIC$ statistic takes the form
\begin{eqnarray}\label{BIC}
BIC(p^*,T)=-2\sum_{n=1}^{N-1}log(f_{\mathbf{R}_n}(\mathbf{r}_n))+log(NT)(Tp^*+2),
\end{eqnarray}
where $\mathbf{r}_n$ is the $T$-dimensional residual vector corresponding to $\mathbf{R}_n$ and $NT$ is the number of observations of the model (\ref{model1}). Note that starting from $n=1$ means that we actually take into account the truncated sequence of residuals $[R_{T+1},\cdots,R_{NT}]$. If $\{Z_t\}$ is a sequence of Gaussian distributed random variables, then the PDF used in Eq. (\ref{BIC}) is given by Eq. (\ref{pdf_R}).  If {the sequence $\{Z_t\}$ is not Gaussian}, we propose to use the methodology based on the inverse CF, see Section \ref{sec:val} for more details.

In case {where} the period $T$ is unknown, we can also apply the $BIC$ statistic defined in Eq. (\ref{BIC}); however, we consider it as a function of two arguments{,} $p^*$ and $T^*$. {We calculate its value for all combinations of $p^*=1,\cdots,p_{max}$ and $T^*=1,\cdots,T_{max}$ such that $p^*<T^*$.} In the final stage, we select the optimal $p_{opt}^{*}$ and $T_{opt}^{*}$ such that they minimize the $BIC$ statistic. The other approach is to apply one of the known methods for {selecting the period $T$} and then apply the above-described methodology assuming that {the} period is known. For other methods of $T$ selection, we refer the readers, for instance, to \cite{broszkiewicz2004detecting,hurd2007periodically}. In the mentioned bibliography positions, the authors discuss efficient methods for period identification for PC models without additive disturbances. However, when the variance of $\{Z_t\}$ is relatively small with respect to the pure model, the classical approaches may also be used. The methodology for identifying the period for the general case will be considered {in future studies}. Let us note that the methodology described above can also be used for the simultaneous order/period identification for the pure PAR model given in Eq. (\ref{parma1}).

\section{Validation of the model - {residual} analysis}\label{sec:val}
In time series analysis, one of the most common techniques for model validation is {residual} analysis. We recall that the residuals of the model {given by Eq. (\ref{res00})} are not independent. {Thus, we examine} the $T$-dimensional residual series corresponding to independent $T$-dimensional random vectors $\mathbf{R}_n$ given by Eq. (\ref{res11}). The first step of the analysis is to confirm whether the $T$-dimensional blocks of the residuals constitute a sample of independent random vectors. For this purpose, we propose to apply one of the tests verifying the independence of multidimensional vectors; see, e.g., \cite{test1,test2,test3}. 
The next step is to test whether the residual series comes from the assumed $T$-dimensional distribution. The situation is relatively easy {if} both the innovation sequence $\{\xi_t\}$ and the additive noise $\{Z_t\}$ are Gaussian distributed{, as then} the residual series is also $T$-dimensional Gaussian. In {this} case, to test the $T$-dimensional Gaussian distribution, we refer the reader to the review of the new developments in affine-invariant tests for multivariate Gaussianity \cite{testy_gauss}. The situation is less standard when the additive noise is not Gaussian{, so that} the residual series is not $T$-dimensional Gaussian as well. To overcome {this} issue, in the following part of this section, we propose a testing procedure that can be used in case of any distribution of the additive noise. It is based on the CF of the residual series, specified in Theorem \ref{th1}.
\begin{theorem}\label{th1}
The characteristic function of $T$-dimensional residual series $\mathbf{R}_n=[R_{nT+1},\cdots,R_{(n+1)T}]$ (see  Eq. (\ref{res00}) and Eq. (\ref{res11})) of the noise-corrupted PAR($p$) model with period $T$, takes the form 
\begin{equation} \label{eq:characteristic_fun}
    \Phi_{\mathbf{R}_n}(t_1,\cdots,t_T)=\prod_{j=1}^{T}\Phi_{\xi}(t_j)\,\prod_{j=1}^{p+T}\Phi_{Z}\left(-\sum_{i=\max\{1,j-p\}}^{\min\{j,T\}}t_i\phi_{p-(j-i)}(i)\right),
\end{equation}
{where the $\Phi_\xi(\cdot)$ and $\Phi_Z(\cdot)$ are the CFs of the $\{\xi_t\}$ and $\{Z_t\}$, respectively, and $\phi_0(k) = -1$ for $k = 1, \cdots, T$.}
\end{theorem}
\begin{proof*}
Using Eq. (\ref{res00}) and the fact that the parameters of the PAR(p) model are periodic with period $T$ we can write that
\begin{align*}
    \Phi_{\mathbf{R}_n}(t_1,\cdots,t_T)=&\mathbb{E}\left[\exp\{it_1R_{nT+1}+it_2R_{nT+2}+\cdots+it_{T}R_{nT+T}\}\right]\\[5pt]=&\mathbb{E}[\exp\{it_1(\xi_{nT+1}+Z_{nT+1}-\phi_1(nT+1)Z_{(nT+1)-1}-\cdots-\phi_{p}(nT+1)Z_{(nT+1)-p})+\\
    &\quad \quad \quad it_2(\xi_{nT+2}+Z_{nT+2}-\phi_1(nT+2)Z_{(nT+2)-1}-\cdots-\phi_{p}(nT+2)Z_{(nT+2)-p})+\\
    &\quad \quad \quad\cdots+\\
    &\quad \quad \quad it_T(\xi_{nT+T}+Z_{nT+T}-\phi_1(nT+T)Z_{(nT+T)-1}-\cdots-\phi_{p}(nT+T)Z_{(nT+T)-p})
    \}]\\[5pt]=&\mathbb{E}[\exp\{it_1(\xi_{nT+1}+Z_{nT+1}-\phi_1(1)Z_{nT}-\cdots-\phi_{p}(1)Z_{nT+1-p})+\\
    &\quad \quad \quad it_2(\xi_{nT+2}+Z_{nT+2}-\phi_1(2)Z_{nT+1}-\cdots-\phi_{p}(2)Z_{nT+2-p})+\\
    &\quad \quad \quad\cdots+\\
    &\quad \quad \quad it_T(\xi_{nT+T}+Z_{nT+T}-\phi_1(T)Z_{nT+T-1}-\cdots-\phi_{p}(nT+T)Z_{nT+T-p})
    \}].
\end{align*}
Let us note that the argument of the exponential function under expected value can be written as a linear combination of $2T+p$ elements, namely $\xi_{nT+1},\xi_{nT+2},\cdots,\xi_{nT+T}$ and $Z_{nT+1-p},Z_{nT+2-p},\cdots,Z_{nT+T}$, which are all independent random variables. Using the above remark and the fact that $$\xi_{nT+1}\stackrel{d}{=}\xi_{nT+2}\stackrel{d}{=}\cdots\stackrel{d}{=}\xi_{nT+T}$$ and
$$Z_{nT+1-p}\stackrel{d}{=}Z_{nT+2-p}\stackrel{d}{=}\cdots\stackrel{d}{=}Z_{nT+T}$$
one can write the formula given in Eq. (\ref{eq:characteristic_fun}). \qed
\end{proof*}
\begin{example}
Let $\{Y_t\}$ denote the PAR model with additive noise defined in Eq. (\ref{model1}) with Gaussian innovations $\{\xi_t\}$ and Gaussian additive noise $\{Z_t\}$. Then, assuming $T=2$ and $p=1$ and using Eq. (\ref{eq:characteristic_fun}), {we obtain the following characteristic function for the residual vector $\mathbf{R}_n=[R_{2n+1},R_{2n+2}]$:}
\begin{align}
    \Phi_{\mathbf{R}_n}(t_1,t_2)=&\exp\left(-\frac{1}{2}\left(t_1^2(\sigma_\xi^2+\phi^2(1)\sigma_Z^2+\sigma_Z^2)-2t_1t_2\phi(2)\sigma_Z^2+t_2^2(\sigma_\xi^2+\phi^2(2)\sigma_Z^2+\sigma_Z^2)\right)\right),
\end{align}
which is simply the characteristic function of {a} two-dimensional zero-mean Gaussian random vector with the elements of the covariance matrix $\boldsymbol{\Gamma}^R=[\gamma^R(k,l)]_{k,l=1}^{2}$ given in Eq. (\ref{gamma}). 
\end{example}
\begin{example} \label{ex:ex2}
Let $\{Y_t\}$ denote the PAR model with additive noise defined in Eq. (\ref{model1}). Moreover, let us assume that the innovations $\{\xi_t\}$ are Gaussian, whereas the additive disturbances $\{Z_t\}$ are mixture of two Gaussian distributions with parameters $a_1,a_2$, $\sigma_1^2,\sigma_2^2$ and $\mu_1=\mu_2=0$, see \ref{appa} for more details. Then, using Eq. (\ref{eq:characteristic_fun}) for $T=2$ and $p=1$, {we obtain the following characteristic function for the residual vector $\mathbf{R}_n=[R_{2n+1},R_{2n+2}]$:}
\begingroup
\allowdisplaybreaks
\begin{align} \label{eq:chfun_ex2}
    \Phi_{\mathbf{R}_n}(t_1,t_2)&=a_1^3\exp\left(-\frac{1}{2}\left(t_1^2(\sigma_\xi^2+\phi^2(1)\sigma_1^2+\sigma_1^2)-2t_1t_2\phi(2)\sigma_1^2+t_2^2(\sigma_\xi^2+\phi^2(2)\sigma_1^2+\sigma_1^2)\right)\right)\\
    &+a_1^2a_2\exp\left(-\frac{1}{2}\left(t_1^2(\sigma_\xi^2+\phi^2(1)\sigma_1^2+\sigma_1^2)-2t_1t_2\phi(2)\sigma_1^2+t_2^2(\sigma_\xi^2+\phi^2(2)\sigma_1^2+\sigma_2^2)\right)\right)\nonumber\\
    &+a_1^2a_2\exp\left(-\frac{1}{2}\left(t_1^2(\sigma_\xi^2+\phi^2(1)\sigma_1^2+\sigma_2^2)-2t_1t_2\phi(2)\sigma_2^2+t_2^2(\sigma_\xi^2+\phi^2(2)\sigma_2^2+\sigma_1^2)\right)\right)\nonumber\\
    &+a_1a_2^2\exp\left(-\frac{1}{2}\left(t_1^2(\sigma_\xi^2+\phi^2(1)\sigma_1^2+\sigma_2^2)-2t_1t_2\phi(2)\sigma_2^2+t_2^2(\sigma_\xi^2+\phi^2(2)\sigma_2^2+\sigma_2^2)\right)\right)\nonumber\\
    &+a_1^2a_2\exp\left(-\frac{1}{2}\left(t_1^2(\sigma_\xi^2+\phi^2(1)\sigma_2^2+\sigma_1^2)-2t_1t_2\phi(2)\sigma_1^2+t_2^2(\sigma_\xi^2+\phi^2(2)\sigma_1^2+\sigma_1^2)\right)\right)\nonumber\\
    &+a_1a_2^2\exp\left(-\frac{1}{2}\left(t_1^2(\sigma_\xi^2+\phi^2(1)\sigma_2^2+\sigma_1^2)-2t_1t_2\phi(2)\sigma_1^2+t_2^2(\sigma_\xi^2+\phi^2(2)\sigma_1^2+\sigma_2^2)\right)\right)\nonumber\\
    &+a_1a_2^2\exp\left(-\frac{1}{2}\left(t_1^2(\sigma_\xi^2+\phi^2(1)\sigma_2^2+\sigma_2^2)-2t_1t_2\phi(2)\sigma_2^2+t_2^2(\sigma_\xi^2+\phi^2(2)\sigma_2^2+\sigma_1^2)\right)\right)\nonumber\\
    &+a_2^3\exp\left(-\frac{1}{2}\left(t_1^2(\sigma_\xi^2+\phi^2(1)\sigma_2^2+\sigma_2^2)-2t_1t_2\phi(2)\sigma_2^2+t_2^2(\sigma_\xi^2+\phi^2(2)\sigma_2^2+\sigma_2^2)\right)\right). \nonumber
\end{align}
\endgroup
{It should be noted that the function above is the CF of a mixture of eight two-dimensional Gaussian distributions. The distribution of the residual vector in this case can also be obtained using the matrix representation  provided in \ref{appaa}. Therefore, the PDF of the residual vector can be expressed as a linear combination of eight zero-mean two-dimensional Gaussian probability density functions. For further explanation, please refer to \ref{appbb}.}
\end{example}

The CF presented in Theorem \ref{th1} can be applied to the residuals distribution testing that is crucial for the validation of the considered model. The procedure is similar to the methodology presented in \cite{Sabuka}, where the authors introduced a testing algorithm based on the distance between the empirical (for a given vector of observations) and theoretical (from tested distribution) CFs. The proposed test is especially useful when the tested distribution does not rely on a closed-form PDF. In the mentioned bibliography position, the methodology was applied {to} the bivariate sub-Gaussian distributed random vectors; however, it is universal and can be extended to any distribution with a closed-form CF, see also \cite{Iskander1}. 

A sketch of the proposed test is as follows. We consider the following test statistic:
\begin{equation} \label{eq:statistics_test}
    D = \sup_{t\in\mathbb{R}^n}\left|\hat{\Phi}_{\mathbf{R}_n}(\mathbf{t})-\Phi_{\mathbf{R}_n}(\mathbf{t})\right|,
\end{equation}
where $\hat{\Phi}_{\mathbf{R}_n}(\cdot)$ and ${\Phi}_{\mathbf{R}_n}(\cdot)$ denote the empirical and theoretical characteristic {functions} of the residual series ${\mathbf{R}_n}$, respectively. Moreover, let us formulate the null hypothesis ($H_0$) that the vector of observations corresponds to the model defined in Eq. (\ref{model1}) with specified distributions of the innovations and additive noise for specified set of all parameters. In practice, those parameters are estimated using the procedures presented in \ref{appB} {and, in consequence, the specification of  the null hypothesis is done based on the parameters obtained from the numerical search.}
The alternative hypothesis ($H_a$) states that the above model does not fit the data considered. The testing procedure can be briefly described as follows: (a) For the data under consideration estimate the parameters of PAR model with additive noise and calculate the test statistic $D$ given in Eq. (\ref{eq:statistics_test}); (b) Assume that the model fitted in (a) is the model under $H_0$; (c) Simulate the data from the model specified in $H_0$; (d) For the simulated data, calculate the residuals corresponding to $\mathbf{R}_n$ under $H_0$; (e) Calculate the test statistic given in Eq. (\ref{eq:statistics_test}) - denote it as $D_1$; (f) Repeat the points (c)-(e) $M$ times to obtain $D_1,D_2,\cdots,D_M$ values; (g) Calculate the p-value as a fraction of $D_i$'s, $i=1,\cdots,M$, that exceed $D$; (h) If the p-value is less that the chosen significance level, reject the $H_0$ hypothesis. 

We mention here that formula (\ref{eq:characteristic_fun}) can be used to find the PDF of the residual series by applying the inverse Fourier transform of CF \cite{ifft}. Obviously, the inverse Fourier transform is calculated numerically. In this paper, we use the multivariate version of {the fast Fourier transform-based algorithm, proposed in \cite{witkovsky}}. The PDF is approximated on the grid consisting of $(2^5)^T$ points, and its value for in-between points is computed using linear interpolation. The {resulting} PDF is further used for the $BIC$ statistic (\ref{BIC}) to obtain the optimal $p$ and $T$ parameters. More details on that aspect are presented in Section \ref{sec:identification}. \footnote{ {All algorithms used in this paper  were prepared in MATLAB R2021b and are available upon the special request of the reader. }}

\section{Simulation study}\label{sec:simul}

\subsection{Identification of the optimal $p$ and $T$ parameters} \label{sec:identification}

In this part, the methodology for the selection of model order $p$ and period $T$ presented in Section \ref{ident} is assessed using Monte Carlo simulations with sample trajectories generated from the noise-corrupted PAR model $\{Y_t\}$ given in Eq. \eqref{model1}. Here, we consider models with $p=1,2,3$ and $T=4$, with the following matrices of PAR coefficients $\Phi^{(p)}$:
\begin{eqnarray}\label{PARexcoefs}
\Phi^{(1)} =
\begin{bmatrix}
-0.1208 \\
-0.5773 \\
-0.0362 \\
-0.3254\\
\end{bmatrix}
\quad
\Phi^{(2)} =
\begin{bmatrix}
-0.1208 & -0.0878 \\
-0.5773 & -0.9798\\
-0.0362 & 0.9196\\
-0.3254 & -0.5802\\
\end{bmatrix}
\quad
\Phi^{(3)} =
\begin{bmatrix}
-0.1208 & -0.0878 & 0.6605\\
-0.5773 & -0.9798 & -0.6826\\
-0.0362 & 0.9196 & 0.6555\\
-0.3254 & -0.5802 & -0.5313\\
\end{bmatrix},
\end{eqnarray}
where the element in the $v$-th row and $i$-th column is the $\phi_i(v)$ coefficient, for $v=1,\cdots,T$ and $i=1,\cdots,p$. In the whole simulation study, we assume standard Gaussian innovations of PAR model $\{\xi_t\} \sim N(0,\sigma_\xi^2=1)$.

\begin{figure}[H]
    \centering
    \includegraphics[width=0.8\textwidth]{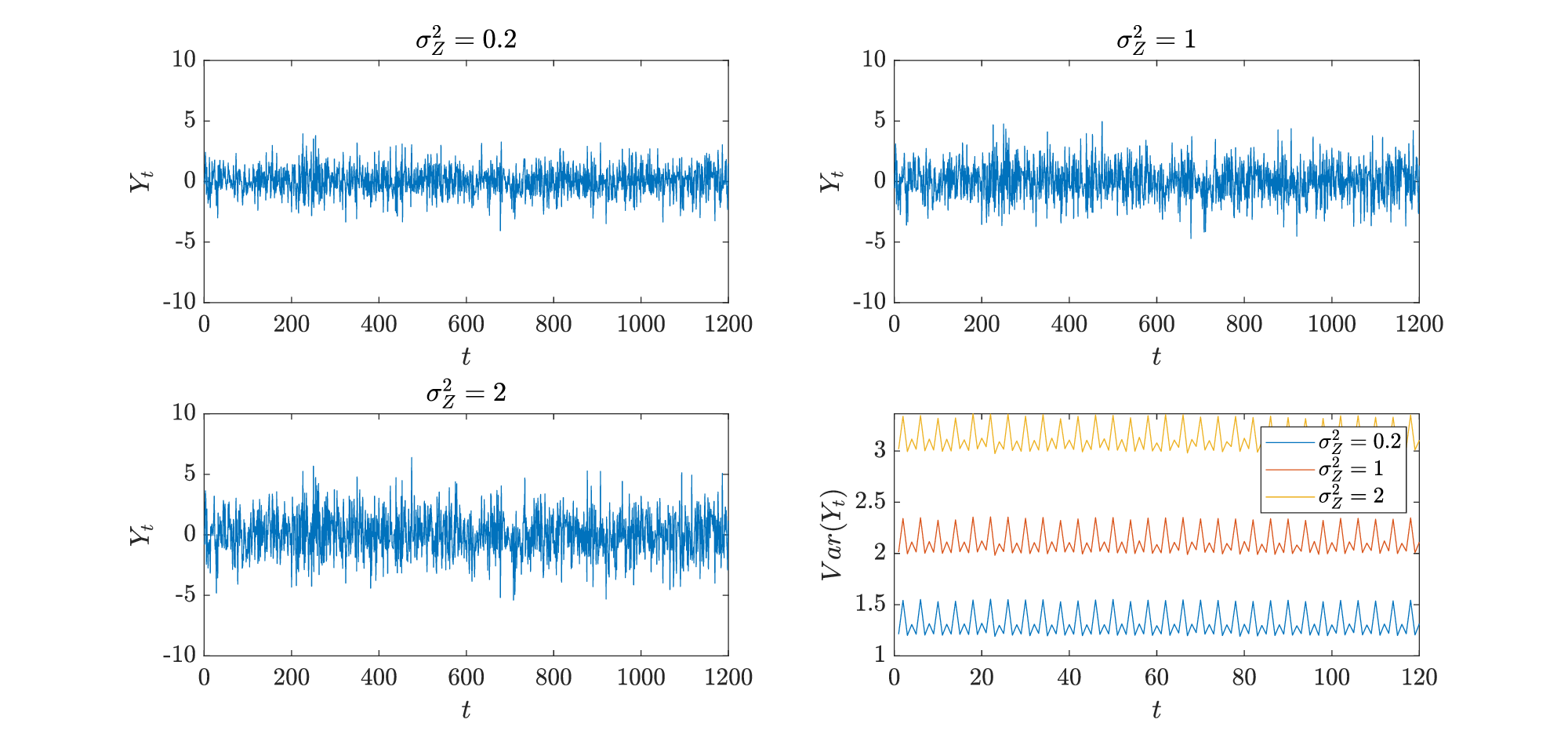}
    \caption{Sample trajectories of $\{Y_t\}$ time series and empirical variances of $Y_t$ for each $t$ calculated from 100000 trajectories for different values of Gaussian additive noise variance in the case of $\Phi^{(1)}$ coefficients ($p=1$, $T=4$, see Eq. \eqref{PARexcoefs}).}
    \label{fig:vars_p1}
\end{figure}

\begin{figure}[H]
    \centering
    \includegraphics[width=0.3\textwidth]{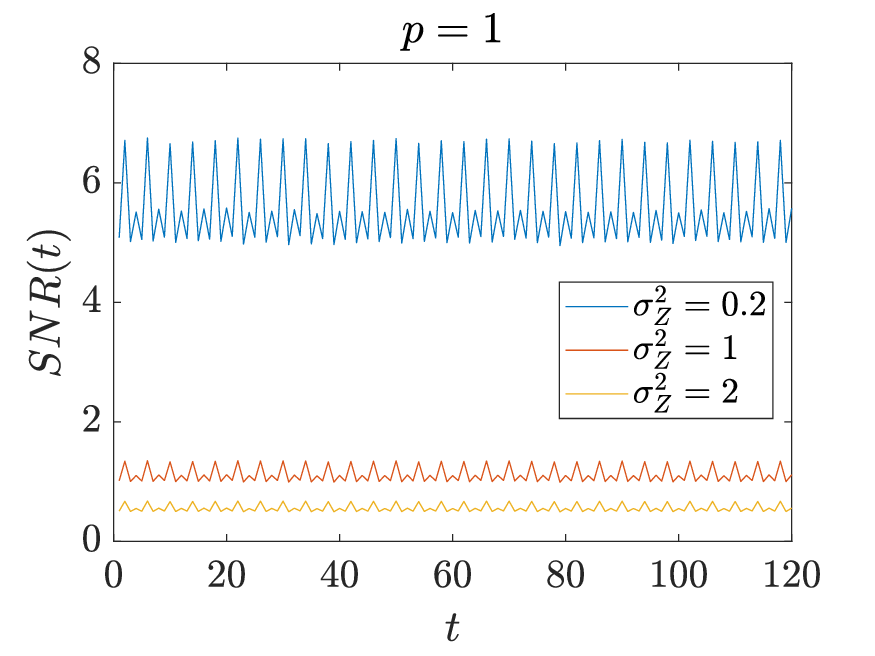}
    \includegraphics[width=0.3\textwidth]{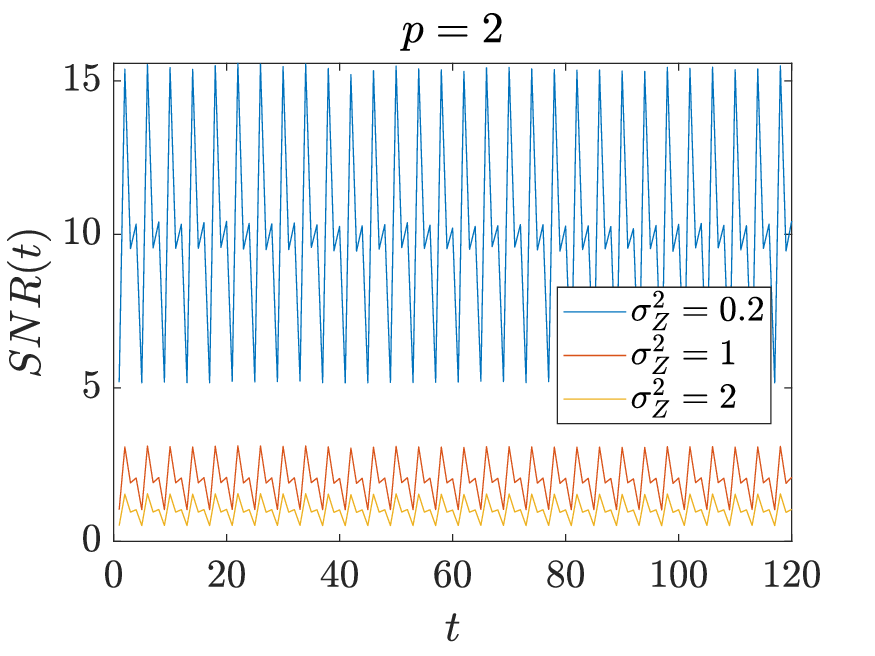}
    \includegraphics[width=0.3\textwidth]{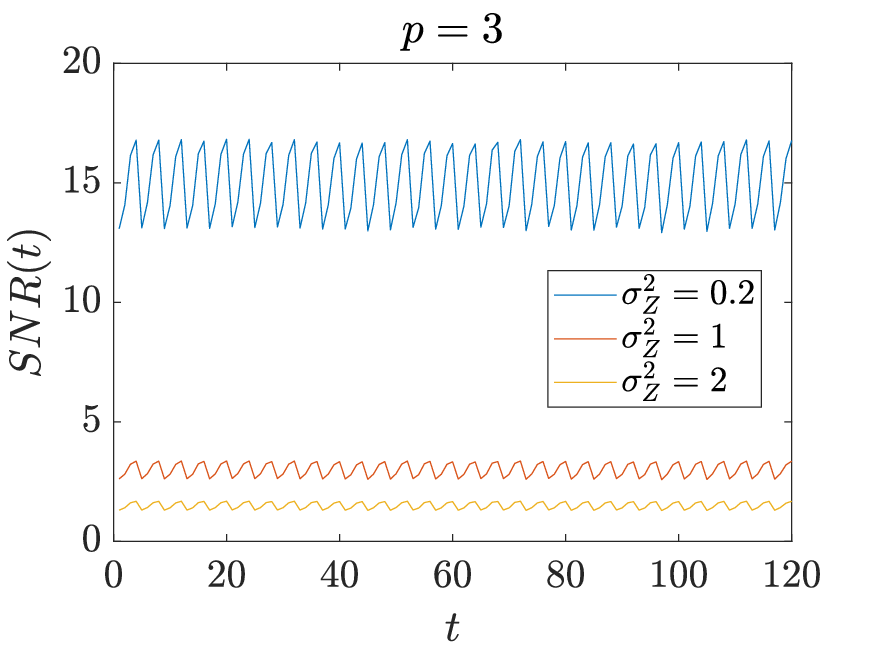}

    \caption{Empirical signal-to-noise ratios (see Eq. (\ref{snr})) calculated for 100000 trajectories of $\{Y_t\}$ time series with coefficients $\Phi^{(p)}$ (see Eq. \eqref{PARexcoefs}) and different values of $\sigma_Z^2$.}
    \label{fig:snr}
\end{figure}

{The first experiment} performed concerns the selection of {the} order $p$ in the situation with the known period $T=4$ and Gaussian-distributed additive noise. The procedure is as follows. We simulate $M=1000$ trajectories of length $NT = 1200$ from the PAR model of order $p$ (i.e. with coefficients $\Phi^{(p)}$, see Eq. \eqref{PARexcoefs}) with additive noise $\{Z_t\} \sim N(0,\sigma_Z^2)$. Then, for each generated trajectory, the $BIC(p^*,T)$ statistic is calculated for $p^*=1,2,3$ (see Eq. (\ref{BIC})). The $p^*_{opt}$ is the order $p^*$ {that} minimizes $BIC(p^*,T)$. In calculations, to obtain residual blocks $\mathbf{r}_n$ for $n=1,\cdots,N-1$, we estimate the coefficients of the PAR model using the modified errors-in-variables method described in \ref{appB}. This experiment is performed for $p=1,\,2,\,3$ and for $\sigma_Z^2=0.2,\,1,\,2$. The time series sample trajectories considered in these simulations are presented in Fig. \ref{fig:vars_p1}. The same sample of pure model $\{X_t\}$ is taken for each $\sigma_Z^2$. Moreover, on the mentioned figure, the empirical variances (calculated from 100000 trajectories) are also plotted. Due to space limitations, we present only the samples for $p=1$, but for $p=2,3$ the observed behaviour is similar.  It can be seen that for increasing additive noise variance, the underlying PAR trajectory becomes more hidden. This is confirmed by empirical signal-to-noise ratios defined {as}
\begin{eqnarray}\label{snr}
SNR(t) = \frac{Var(X_t)}{Var(Z_t)},
\end{eqnarray}
calculated for 100000 trajectories and illustrated in Fig. \ref{fig:snr}, now for each $p=1,2,3$. Because with larger $\sigma_Z^2$ the pure PAR process is more disturbed, the main challenge {here} is to provide reliable results also for low signal-to-noise ratio samples.

\begin{figure}
    \centering
    \includegraphics[width=\textwidth]{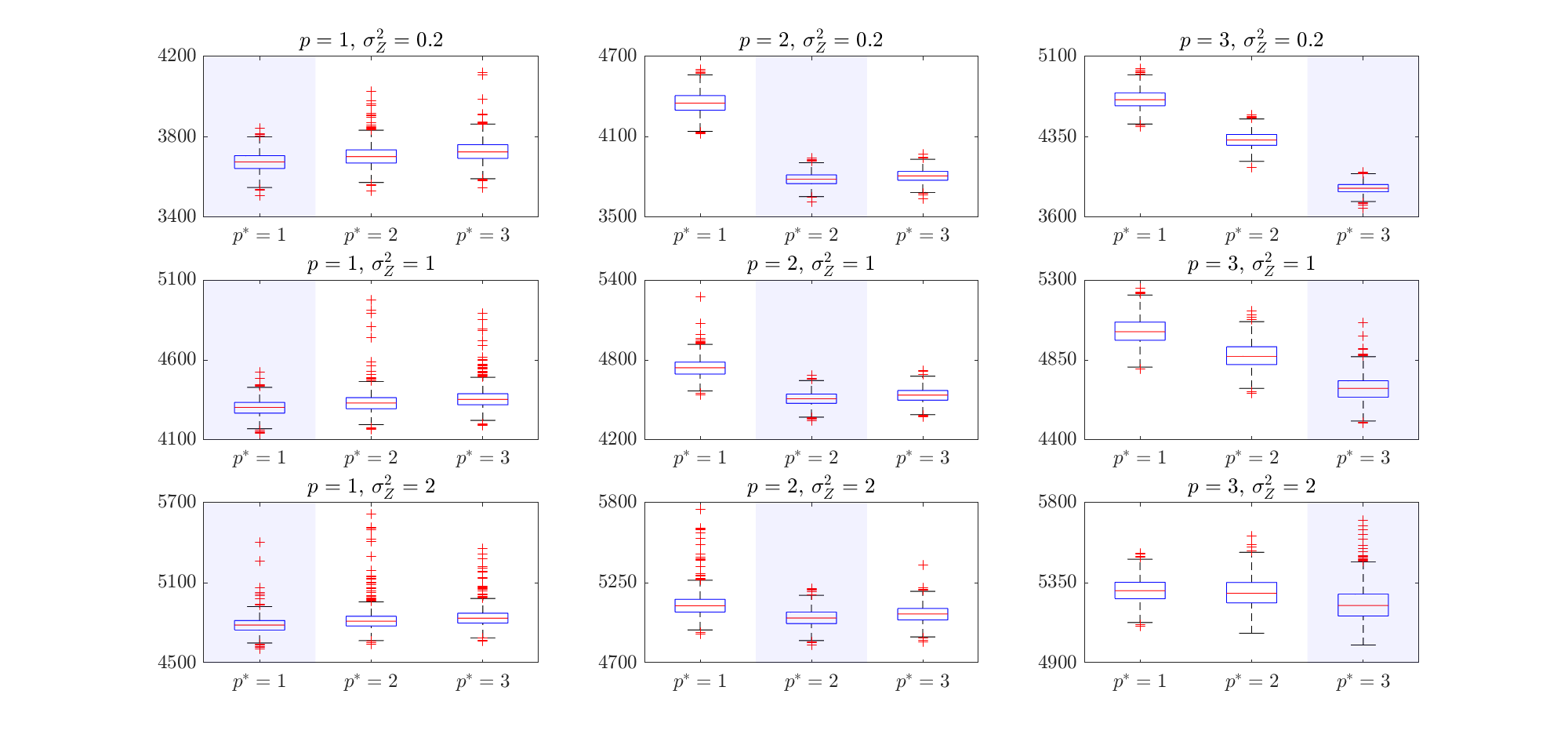}
    \caption{Boxplots of $BIC(p^*,T)$ values for checked orders $p^*=1,2,3$ calculated in case of known $T=4$ for true model orders $p=1,2,3$ and $\{Z_t\} \sim N(0,\sigma_Z^2)$ with $\sigma_Z^2=0.2,1,2$. The shaded area on each subplot refers to the boxplot for $p^*=p$.}
    \label{fig:p_gauss}
\end{figure}

The boxplots of all $BIC$ values in these simulations set are presented in Fig. \ref{fig:p_gauss}. Let us first analyze the results from the top panel, i.e. for the lowest considered additive noise variance $\sigma_Z^2=0.2$. One can see that for each $p$ the results for $p^*<p$ (if any) are visibly larger than the remaining ones. Furthermore, starting from $p^*=p$, the values of $BIC$ seem to stabilize. Both of these characteristics  are present in the case of $p=2$. This behaviour is caused by the fact that for $p^*>p$ the model can also be adjusted properly, so that the estimated coefficients $\{\phi_i(v)\}$ for $i=1,\cdots,p$ are close to the true ones and for $i=p+1,\cdots,p^*$ are nearly zero. 
From the presented results, one can conclude that instead of the $BIC$ minimization, one can consider another approach of order identification, i.e., detecting the order $p^*$ for which the described stabilization behaviour starts to occur. For the case of $\sigma_Z^2=1$ (middle panel), although the results for all $p^*$ are now relatively closer to each other, one can see a pattern similar to that before. In particular, one can still distinguish between the results for $p^*<p$ and $p^* \geq p$. However, when additive noise variance is even larger (see the bottom panel of Fig. \ref{fig:p_gauss}, where $\sigma_Z^2=2$), the difference between boxplots for different $p^*$ becomes much less visible. Hence, in this case, the order identification {becomes more challenging, as expected,} considering the lower signal-to-noise ratio. 
We also calculate the fraction of correctly identified orders by $BIC$ minimization, i.e. the percentage of cases {where} $p^*_{opt}=p$, for each considered combination of $p$ and $\sigma_Z^2$. The results are presented in {Table} \ref{tab:fracs}. Most of all, we can see that the identification efficiency indeed drops for a larger additive noise variance. However, in general, the results show that in the vast majority of cases the proposed criterion managed to identify the order correctly, even for $\sigma_Z^2=2$. Let us recall that the order selection here was done by the minimization of $BIC$, despite the stabilization behaviour mentioned above, which could result in a wrong selection (i.e., $BIC$ would tend to choose larger order than the correct one).
\renewcommand{\arraystretch}{1.1}
\begin{table}
    \centering
    \begin{tabular}{|c|c|c|c||c|c|c|}
    \hline
    & \multicolumn{3}{c||}{$\{Z_t\} \sim N(0,\sigma_Z^2)$} & \multicolumn{3}{c|}{$\{Z_t\} = \sigma_Z \cdot \{\tilde{Z}_t\}$} \\\hline
         & $p=1$ & $p=2$ & $p=3$ & $p=1$ & $p=2$ & $p=3$  \\ \hline \hline
         $\sigma_Z^2 = 0.2$ & 98.7\%  & 99.9\% & 100\% & 98.5\%  & 100\% & 100\%\\ \hline
         $\sigma_Z^2 = 1$ & 96\%  & 96.8\% & 99.2\% & 93.9\%  & 97.8\% & 99.2\%  \\ \hline
          $\sigma_Z^2 = 2$ & 91.8\% & 86.1\% & 73.0\% & 90.2\% & 85.8\% & 76.6\% \\ \hline
    \end{tabular}
    \caption{Fraction of correctly identified orders in case of known $T=4$ for different $p$, $\sigma_Z^2$ and both considered additive noise distributions. Here  $\{\tilde{Z}_t\}$ is a sequence of  mixture Gaussian distributed random variables with $a_1=a_2=0.5$, $\mu_1=\mu_2=0$ and $\sigma_1^2=0.5$, $\sigma_2^2=1.5$.}
    \label{tab:fracs}
\end{table}

Next, let us present the same experiment as the one described above but for another type of additive noise distribution. Let us consider a sequence $\{\tilde{Z}_t\}$ of i.i.d. random variables from {a} mixture of $m=2$ Gaussian distributions (see \ref{appa}) with $a_1=a_2=0.5$, $\mu_1=\mu_2=0$ and $\sigma_1^2=0.5$, $\sigma_2^2=1.5$. These parameters are tuned in such a way that $\tilde{Z}_t$ for each $t$ has a unit variance. Furthermore, this mixture yields a heavy-tailed distribution as its excess kurtosis (see \ref{appa}) is positive. As the additive noise sequences used in simulations described below we use scaled versions of $\{\tilde{Z}_t\}$, setting $Z_t = \sigma_Z \cdot \tilde{Z}_t$, again taking $\sigma_Z^2=0.2,1,2$. For signal-to-noise ratios for this case, see again Fig. \ref{snr}.
{In this case, to present the proposed methodology,  we use the procedure presented in Section \ref{sec:val} based on the CF inversion.} {However, it should be noted that   the procedure of inverting the CF  could be ignored here, since the PDF of the residuals is given in the explicit form, see \ref{appbb} for more details. In this part, the whole procedure is presented for illustration of the introduced methodology for the case where the additive noise has different than the Gaussian distribution. } 

Let us note that in this case the estimation stage provides information about $\sigma_Z^2$ (in the form of its estimate $\hat{\sigma}_Z^2$), but not about the structure of the mixture (i.e., the form of $\tilde{Z}_t$) that is also needed in the calculation of the CF. Hence, here we assume that this structure is known and only the additive noise variance may vary dependent on estimation results. In other words, during the CF computation for a trajectory with estimated additive noise variance $\hat{\sigma}_Z^2$, we assume that $Z_t = \hat{\sigma}_Z \cdot \tilde{Z}_t$. 

\begin{figure}
    \centering
    \includegraphics[width=\textwidth]{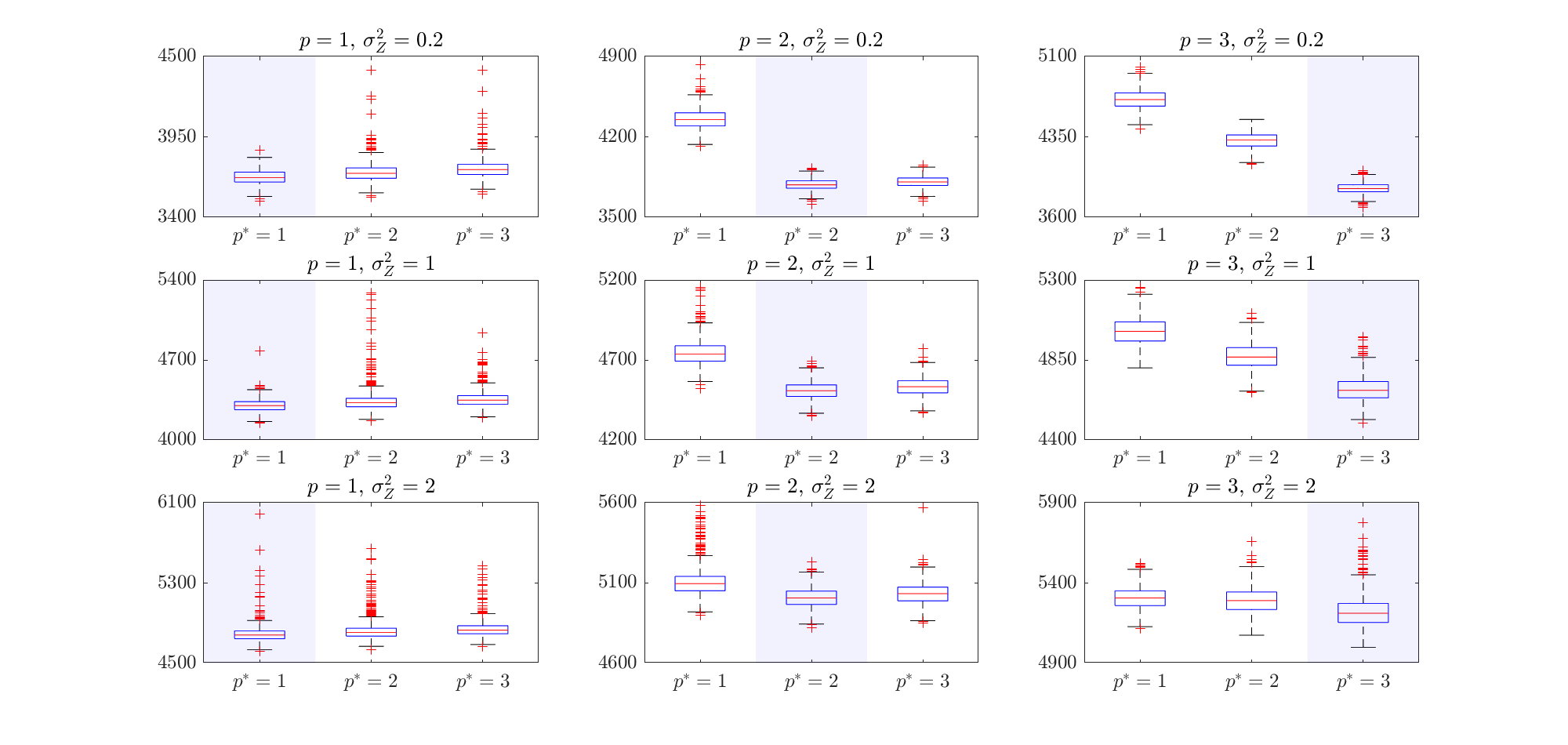}
    \caption{Boxplots of $BIC(p^*,T)$ values for checked orders $p^*=1,2,3$ calculated in case of known $T=4$ for true model orders $p=1,2,3$ and $\{Z_t\} = \sigma_Z \cdot \{\tilde{Z}_t\}$ with $\sigma_Z^2=0.2,1,2$, where  $\{\tilde{Z}_t\}$ is a sequence of mixture Gaussian distributed random variables with $a_1=a_2=0.5$, $\mu_1=\mu_2=0$ and $\sigma_1^2=0.5$, $\sigma_2^2=1.5$. The shaded area on each subplot refers to the boxplot for $p^*=p$.}
    \label{fig:p_mixgauss}
\end{figure}

The $BIC$ statistic calculated for trajectories with additive noise from mixture Gaussian distribution (all other details of these simulations are the same as for the Gaussian case) are presented in Fig. \ref{fig:p_mixgauss}. One can see that observed characteristics of the results are similar to those obtained for the Gaussian case described above. Most {importantly}, once again the aforementioned stabilization for $p^* \geq p$ occurs, and the results for different $p^*$ become less distinguishable with increasing additive noise variance. The latter can be again seen in fractions of correctly identified orders which are presented in {Table} \ref{tab:fracs}. Let us note that these results are at a similar level as the proportions obtained for samples with Gaussian additive noise, even though here we only use an approximation of {the} PDF of the residuals. Hence, one can see that the presented approach shows an acceptable efficiency even for non-Gaussian cases.

\begin{figure}
    \centering
    \includegraphics[width=\textwidth]{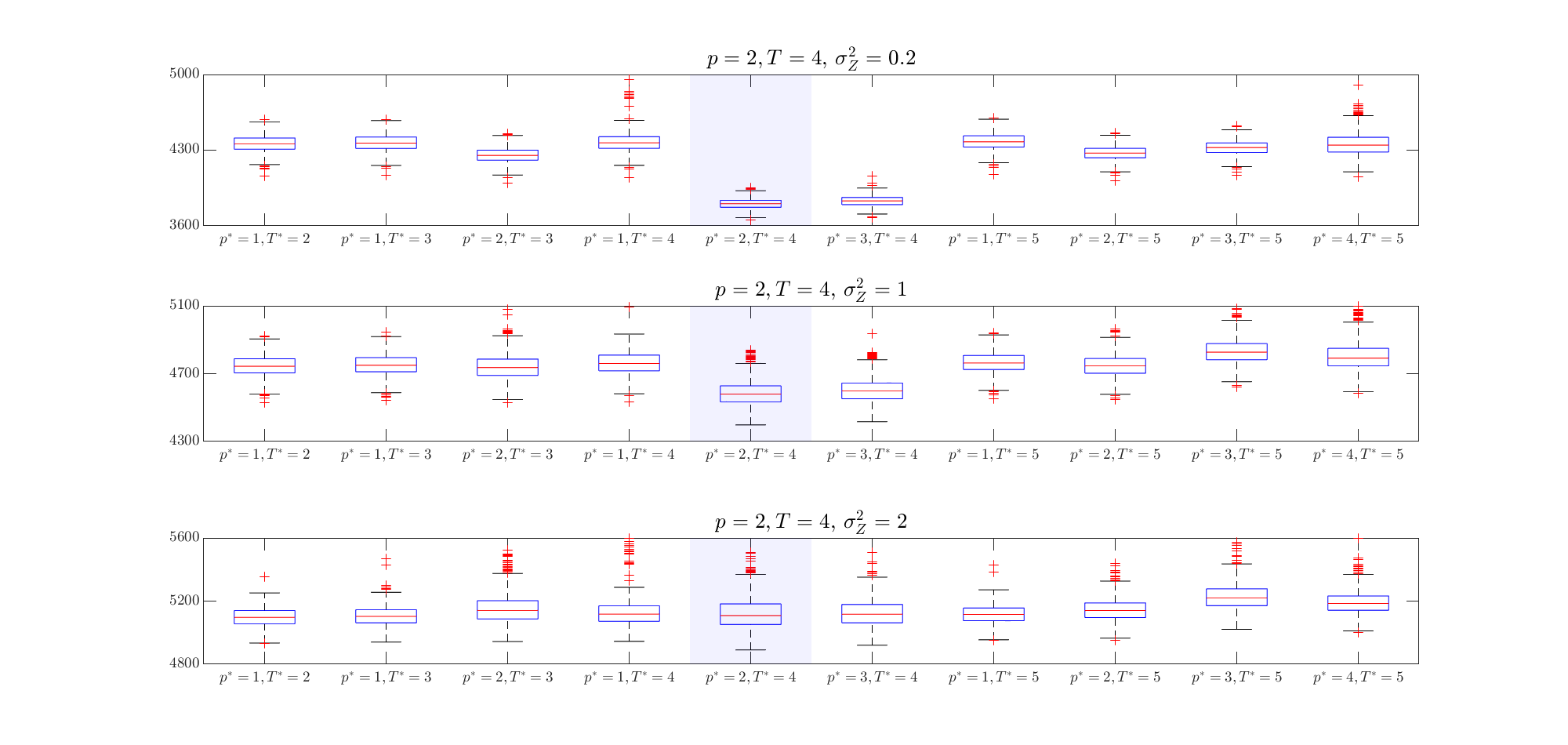}
    \caption{Boxplots of $BIC(p^*,T^*)$ values for all checked pairs $p^*,T^*$ calculated for $\{Z_t\} \sim N(0,\sigma_Z^2)$ with $\sigma_Z^2=0.2,1,2$ in case of unknown $T$. The shaded area on each subplot refers to the boxplot for $p^*=p=2$, $T^*=T=4$.}
    \label{fig:Tp_gauss}
\end{figure}

In the last part of the simulation study, let us consider the problem of identification of both{, the order $p$ and the period $T$, simultaneously,} using the $BIC(p^*,T^*)$ statistic. Here, we only analyze the case with Gaussian additive noise. The simulations procedure performed here, similar to the ones presented before, is as follows. We generate $M=1000$ trajectories of length $1205$ of the PAR model with $\Phi^{(2)}$ coefficients (thus, $p=2$ and $T=4$, see Eq. \eqref{PARexcoefs}) and additive noise $\{Z_t\} \sim N(0,\sigma_Z^2)$. Next, for each simulated sample path, we calculate the $BIC(p^*,T^*)$ statistic for all pairs $p^*,T^* \in \{1,2,3,4,5\}$ such that $p^*<T^*$. We also identify $p^*_{opt}$ and $T^*_{opt}$, i.e. the pair that minimizes the computed criterion. Such choice of sample length  is made to ensure that for all checked potential periods the sequence of residuals considered in $BIC$ calculation has the same length. Here, for each $T^*$, we use the sequence $[R_{T^*_{max}+1},\cdots,R_L]$, where $T^*_{max}=5$, which here is of length 1200 (note that it is divisible by each considered $T^*$). As before, we consider three levels of additive noise variance, namely $\sigma_Z^2=0.2,1,2$.
The calculated $BIC$ results are illustrated in Fig. \ref{fig:Tp_gauss}. First, let us analyze the case where $\sigma_Z^2=0.2$. One can clearly see that the lowest values were obtained for $p^*=2$ and $T^*=4$, which is {the} actual order/period pair of the simulated model, and for $p^*=3$, $T^*=4$, so for the true period and the order larger than the true one. Let us note that this is consistent with the stabilization behaviour observed in {the case of known $T$}. Moreover, since the results for $T^*\neq T$ are at a significantly high level so that the $BIC$ statistic would not select them as the period of a model, one can assume that this criterion can serve as a period identification method for noise-corrupted PAR models. However, {as before}, for larger values of additive noise variance, the $BIC$ results for all considered $p^*,T^*$ pairs become relatively closer to each other so that the proper detection of order and period is then more challenging. Although for $\sigma_Z^2=1$ the $p^*=2$, $T^*=4$ pair still stands out, for $\sigma_Z^2=2$ all boxplots are in fact at a similar level.
We calculate the fraction of correctly identified values of $p$ and $T$, i.e. the percentage of cases where both $p^*_{opt}=p$ and $T^*_{opt}=T$. Moreover, we also consider the proportion of trajectories for which just the latter equality is satisfied (regardless of the selected order). Let us note that this result indicates the performance of the presented methodology in period detection. The percentages considered for different values $\sigma_Z^2$ are presented in {Table} \ref{tab:Tp_gauss}. {As before}, the model identification efficiency drops with increasing additive noise variance from an almost perfect performance for $\sigma_Z^2=0.2$ to a significantly worse one for $\sigma_Z^2=2$. This can be seen for both simultaneous selections of $p$ and $T$ selection as well as for just the identification of the period. However, let us note that the considered $BIC$ statistic may sometimes be able to find the correct $T$ even if the order obtained is wrong. This is most clearly seen on the difference between both obtained fractions for $\sigma_Z^2=1$. In this case, even though the performance of both $p,T$ selection significantly decreases in comparison to the $\sigma_Z^2=0.2$ situation, for the period identification it is still at a very high level.

\begin{table}
    \centering
    \begin{tabular}{|c|c|c|}
    \hline
         &  $p^*_{opt}=p$, $T^*_{opt}=T$ & $T^*_{opt}=T$\\ \hline \hline
         $\sigma_Z^2 = 0.2$ & 98.9\% & 100\% \\ \hline
         $\sigma_Z^2 = 1$ & 74.1\% & 98.6\% \\ \hline
          $\sigma_Z^2 = 2$ & 37.5\% & 50.7\% \\ \hline
    \end{tabular}
    \caption{Fraction of correctly identified parameters for different $\sigma_Z^2$ with Gaussian additive noise $\{Z_t\}$ in case of unknown $T$.}
    \label{tab:Tp_gauss}
\end{table}

\subsection{Testing {the distribution of residuals}} \label{sec:testing}

\begin{figure}[h!]
    \centering
    \includegraphics[width=0.3\textwidth]{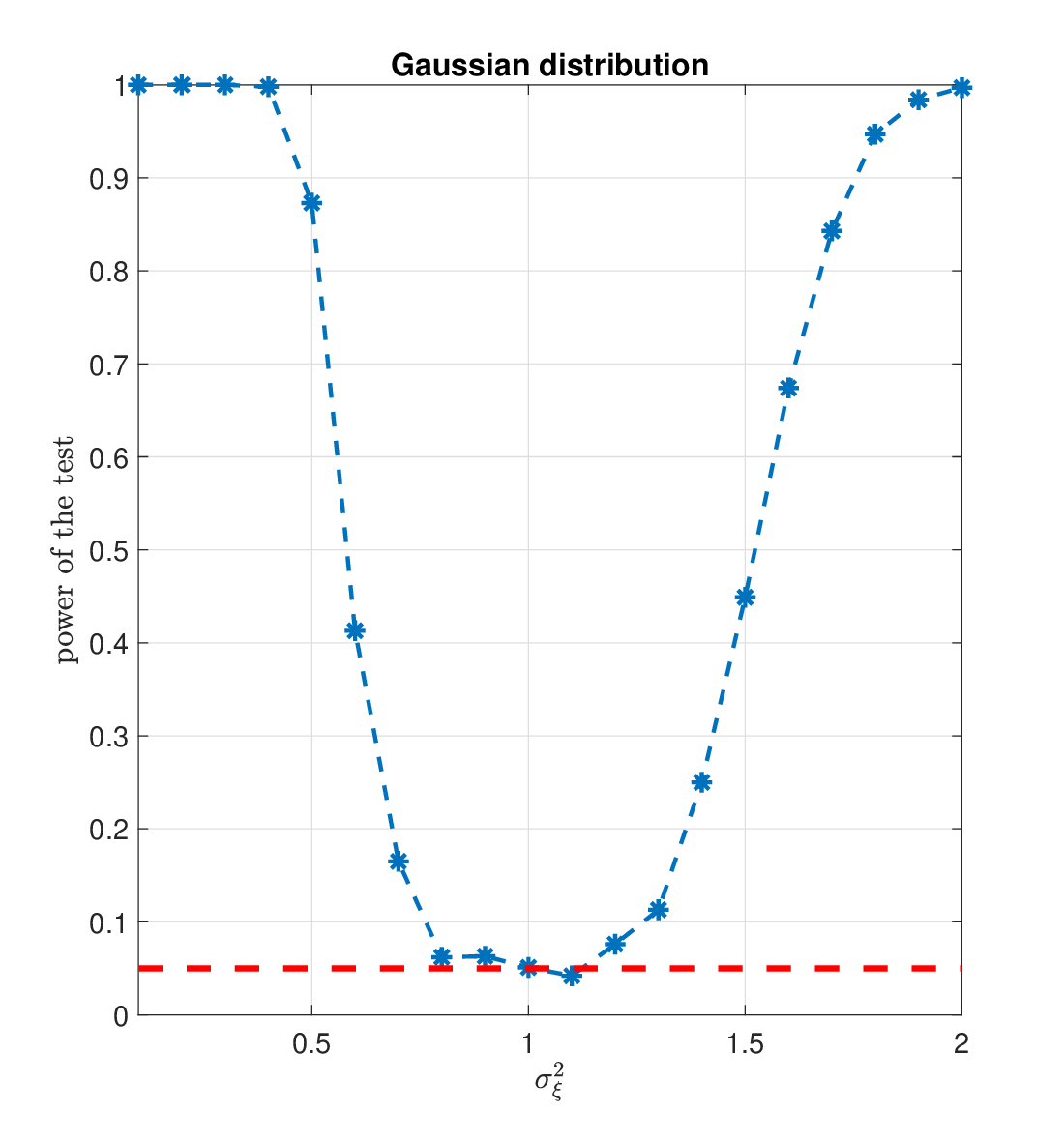}
    \includegraphics[width=0.3\textwidth]{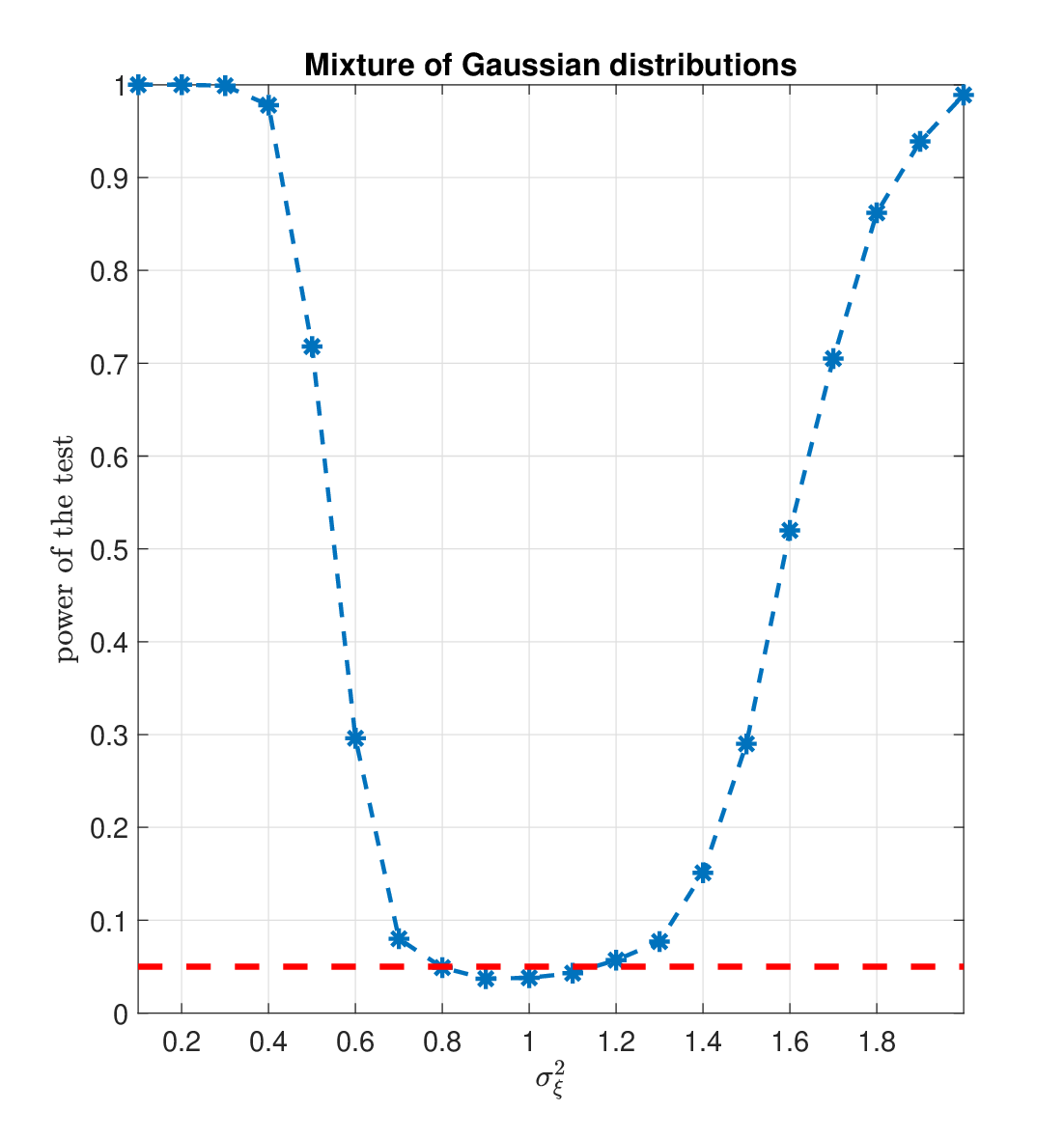}
    
    \caption{The power of the test for different levels of additive noise variance. In $H_0$ we assume PAR(1) model with $T=2$ ($\phi(1)=0.4$, $\phi(2)=-0.6$), Gaussian innovations ($\sigma^2_\xi$=1) and the additive noise on the left panel is Gaussian ($\sigma_Z^2=1$) and on the right panel is the mixture of two Gaussian distributions ($a_1=a_2=0.5$, $\mu_1=\mu_2=0$, $\sigma_1^2=0.5$, and $\sigma_2^2=1.5$).}
    \label{fig:power1}
\end{figure}

In this part, we present the results of the residual series testing methodology outlined in detail in Section \ref{sec:val}. In the simulations conducted, we focus on assessing the test power, which is considered here as a function of the additive noise variance. Below, we provide a brief overview of the verification process carried out in this section.
The procedure is as follows: under the null hypothesis $H_0$, we assume a certain PAR model of order $1$ with $T=2$, zero-mean, unit variance Gaussian innovations, and a certain type of additive noise. We then simulate $M=1000$ trajectories from that model for different levels of additive noise variance (not necessarily equal to the level assumed in $H_0$, but the other parameters do not change). For these data, we perform the proposed test and calculate the percentage of repetitions where $H_0$ is rejected. The above-mentioned simulations are conducted for a PAR(1) model with $T=2$ ($\phi(1)=0.4$, $\phi(2)=-0.6$), Gaussian innovations ($\sigma^2_\xi=1$), and two types of additive noise distributions: Gaussian ($\sigma_Z^2=1$ for $H_0$) and a mixture of two Gaussians ($a_1=a_2=0.5$, $\mu_1=\mu_2=0$, $\sigma_1^2=0.5$, and $\sigma_2^2=1.5$ for $H_0$).
The results are shown in Fig. \ref{fig:power1}, where on the $y$-axis we present the power of the test for different values of the additive noise variance given on the $x$-axis. It should be noted that $\sigma^2_\xi=1$ corresponds to the null hypothesis $H_0$, and in this case, the power of the test is close to the assumed significance level (here equal to $0.05$). As can be seen, the test power takes slightly higher values for the same level of additive noise variance in the Gaussian case than for the mixture of Gaussian distributions. The test has very low power when the variance of the additional disturbances is close to the one assumed in $H_0$. We can see that the power of the test is close to the significance level when the variance of the additional noise differs by a maximum of $\pm 0.2$ (for the Gaussian) or $\pm 0.3$ (for the mixture Gaussian) from the variance assumed in $H_0$. {The power of the test increases as the distance of the additional noise variance from the level assumed in $H_0$ becomes greater}. However, we can see that the test power functions shown in Fig. \ref{fig:power1} are not symmetrical around 1, and the power is greater for the smaller values of the actual variance of the additional disturbances, so we conclude that it is easier to recognize the wrong model in this case. {It is important to note that the results presented in this section are calculated using a bivariate argument $t$ for the test statistic, as defined by Eq. (\ref{eq:statistics_test}), that was chosen from a square grid of values ranging from $-10$ to $10$ in both dimensions with a step size of $0.01$.}

\section{Conclusions}
In this article, we present a new framework for identifying and validating the noise-corrupted PAR model {where} the data are a mixture of pure PAR time series and additive noise. This scenario has been observed in some real-world data, particularly in the context of machine condition monitoring. We have focused on the procedures for selecting the model order and estimating the period for the assumed model of observation. Finally, a novel method for analyzing residuals is presented. The proposed methodology is universal because it is independent of the distribution of the innovation series of the PAR model and the noise distribution. The efficiency is demonstrated for two types {of} additive noise: Gaussian and {finite mixtures of Gaussian distributions}. The common feature of the introduced methods is the form of the CF of the residual series (presented here as a $T$-dimensional random vector). We believe that the proposed procedures can be applied to various data with expected periodically non-stationary characteristics that are naturally affected by external sources.
\section*{Acknowledgments}
The work of WŻ, RZ and AW is supported by National Center of Science under Sheng2 project No.  UMO-2021/40/Q/ST8/00024 "NonGauMech - New methods of processing non-stationary signals (identification, segmentation, extraction, modeling) with non-Gaussian characteristics for the purpose of monitoring complex mechanical structures". 
\section*{Conflicts of interest}
Declarations of interest: none
\bibliography{mybibliography}

\appendix
\section{{Matrix representation of residuals for noise-corrupted PAR($p$) model}}\label{appaa}

{In this part, the matrix representation of the residual series defined in Eq. \eqref{res11} is provided. Its form is as follows:
    \begin{equation}\label{mat_rep}
        \mathbf{R}_n = \bm{\xi}_n + \mathbf{Z}_n\mathbf{A},
    \end{equation}
    where $\bm{\xi}_n = [\xi_{nT+1},\ldots,\xi_{nT+T}]$ is a $1 \times T$ vector of i.i.d. zero mean  components with variance $\sigma_\xi^2$, $\mathbf{Z}_n = [Z_{nT+T},\ldots,Z_{nT+1-p}]$ is a $1 \times (p+T)$ vector of i.i.d. zero mean components with variance $\sigma_Z^2$, independent of $\bm{\xi}_n$, and $\mathbf{A}$ is a $(p+T) \times T$ matrix  of elements $a_{kl}$ defined as
    \begin{equation}\label{a_elements}
    a_{kl} = -\phi_{k+l-T-1}(l),\quad k=1,\cdots,p+T,\,\,\,l = 1,\cdots,T
    \end{equation}
  with $\phi_0(\cdot)=-1$ and $\phi_j(\cdot) = 0$ for $j\notin \{0,\cdots,p\}$. Let us note that the representation given in Eq. \eqref{mat_rep} immediately produces the covariance matrix of $\mathbf{R}_n$ to be on the form 
    \begin{equation}\label{eq3}
        \mathbf{\Gamma}^R  = \sigma_\xi^2 \mathbf{I}_T + \sigma_Z^2 \mathbf{A'A},
    \end{equation}
    where $\mathbf{I}_T$ is a $T \times T$ identity matrix, leading directly to the formulas in Eq. \eqref{gamma}. }

\section{Mixture of Gaussian distributions}\label{appa}
Let $a_1,a_2,\cdots,a_m$ denote a series of non-negative weights satisfying $\sum_{i=1}^ma_i=1$. Let $F_1(\cdot),F_2(\cdot),\cdots,F_m(\cdot)$ denote an arbitrary sequence of Gaussian cumulative distribution functions (CDFs) and {let} $f_1(\cdot),f_2(\cdot),\cdots,f_m(\cdot)$ {be the PDFs} with means $\mu_1,\mu_2,\cdots,\mu_m$ and variances $\sigma_1^2,\sigma_2^2,\cdots,\sigma_m^2$. A random variable $X$ with the following CDF and PDF:
\begin{eqnarray}\label{mG}
F(z)=\sum_{i=1}^ma_iF_i(z),~~ f(z)=\sum_{i=1}^ma_if_i(z)
\end{eqnarray}
is called a mixture of Gaussian distributions.  In this paper for the distribution of the additive noise  we consider two types of mixture of Gaussian distributions, namely with $m=1$ and $m=2$. In both cases{, the considered distributions are centered}. For $m=1$ this distribution reduces to the  Gaussian distribution. In that case the excess kurtosis is equal to zero. For $m=2$ and $\mu_1=\mu_2=0$ the excess kurtosis is equal to $
Kur=\frac{3(a_1\sigma_1^4+a_2\sigma_2^4)}{(a_1\sigma_1^2+a_2\sigma_2^2)^2}-3$. 

\section{{Distribution of the residuals for PAR($p$) model disturbed by noise with mixture of two Gaussian distributions}}\label{appbb}

{To show that the distribution of the residual vector in Example \ref{ex:ex2} is a mixture of eight zero-mean bivariate Gaussian distributions, one can use the matrix representation given in \ref{appaa}. Let us notice that in that case using Eq. (\ref{mat_rep}) we have
\begin{equation}\label{eq4}
    \mathbf{R}_2 \overset{d}{=} \bm{\xi} + \mathbf{ZA},
\end{equation}
where $\bm{\xi} = [\xi_1,\xi_2]$ is a $1 \times 2$ vector of i.i.d. zero mean Gaussian components with variance $\sigma_\xi^2$, $\mathbf{Z} = [Z_1,Z_2,Z_3]$ is a $1 \times 3$ vector of i.i.d. zero mean components with variance $\sigma_Z^2$, independent of $\bm{\xi}$, and $\mathbf{A}$ is a $3 \times 2$ matrix of constant coefficients of the form 
        \begin{equation}\label{eq2}
        \mathbf{A} = \begin{pmatrix}
0 & 1 \\
1 & -\phi_1(2) \\
-\phi_1(1) & 0
\end{pmatrix}.
    \end{equation}
Now, since the $\{Z_i\}$ are two-component mixtures of Gaussian distributions, they can be represented as
\begin{equation}\label{eq5}
    Z_i \overset{d}{=} I_i X_1^{(i)} + (1-I_i)X_2^{(i)}, \quad i=1,2,3,
\end{equation}
where the $\{X_1^{(i)}\}$ are i.i.d. zero-mean Gaussian variables with variance $\sigma_1^2$, the $\{X_2^{(i)}\}$ are i.i.d. zero-mean Gaussian variables with variance $\sigma_2^2$, and the $\{I_i\}$ are i.i.d. Bernoulli variables with $\mathbb{P}(I_i=1) = a_1$ and $\mathbb{P}(I_i=0) = a_2 = 1-a_1$, with all the variables in Eq. (\ref{eq5}) mutually independent. One can note that there are eight different combinations of the values of $I_i$, and conditionally on these values, the vector $\mathbf{Z}=[Z_1,Z_2,Z_3]$ in Eq. (\ref{eq4}) is zero-mean multivariate Gaussian, with the diagonal covariance matrix $\bm{\Omega} = \text{diag}(\omega_1,\omega_2,\omega_3)$ where
\begin{equation}\label{eq6}
    \omega_i = I_i\sigma_1^2 + (1-I_i)\sigma_2^2, \quad i=1,2,3.
\end{equation}
Thus, conditionally on the values of $I_i$, the random vector $\mathbf{R}_2$ in Eq. (\ref{eq4}) is also multivariate Gaussian, with the following covariance matrix:
\begin{equation}\label{eq7}
    \sigma_\xi^2\mathbf{I}_2 + \mathbf{A}'\bm{\Omega}\mathbf{A}.
\end{equation}
This shows that the distribution of $\mathbf{R}_2$ is a mixture of eight zero-mean bivariate Gaussian distributions, with mixing probabilities given by $\pi_{ijk} = \mathbb{P}(I_1 = i,\, I_2 = j,\, I_3 = k),\, i,\,j,\,k = 1,2$. One can notice that these probabilities reduce to $a_1^3,\, a_1^2 a_2,\, a_1 a_2^2$, and $a_2^3$, used in Eq. (\ref{eq:chfun_ex2}) provided in Example \ref{ex:ex2}. It is also worth mentioning that the above findings could be written in a more general case, assuming that the additional noise is an $m$-component mixture of zero-mean Gaussian distributions with variances $\sigma_1^2,\cdots,\sigma_m^2$. However, as the above example is used only for the purpose of illustrating the general methodology, we choose to omit that part of the presentation.} 

\section{Modified errors-in-variables method of estimation for PAR model with additive noise}\label{appB}

In this part we describe the estimation method  used in this paper. It can be considered as {a} generalization of the algorithm proposed in \cite{diversi1} for autoregressive model with additive noise, utilizing both low- and high-order Yule-Walker equations based on the periodic autocovariance function ${\gamma}^Y(w,k) = \mathbb{E}Y_{nT+w}Y_{nT+w-k}$ (of a zero-mean process). For the process $\{Y_t\}$, for $v=1,\cdots,T$, we define low-order Yule-Walker equations as
\begin{eqnarray}\label{y_loyw}
(\mathbf{\Gamma}^Y_v-\sigma^2_Z \mathbf{I}_p) \mathbf{\Phi}_v = \mathbf{\gamma}^Y_v,
\end{eqnarray}
where 
\begin{eqnarray}\label{x_loyw_def1}
(\mathbf{\Gamma}^Y_v)_{i,j} = \gamma^Y(v-i,j-i),~
\mathbf{\gamma}^Y_v = [\gamma^Y(v,1),\cdots,\gamma^Y(v,p)]',~~
\mathbf{\Phi}_v = [\phi_1(v),\cdots,\phi_p(v)]'
\end{eqnarray}
and $i,j=1,2,\cdots,p$. On the other hand, we define the system of $s$ high-order Yule-Walker equations for the process $\{Y_t\}$, again for $v=1,\cdots,T$, in the following way:
\begin{eqnarray}\label{y_hoyw}
_s\tilde{\mathbf{\Gamma}}^{Y}_v \mathbf{\Phi}_v = {}_s\tilde{\mathbf{\gamma}}^{Y}_v.
\end{eqnarray}
where $_s\tilde{\mathbf{\Gamma}}^{Y}_v$ is a $s \times p$ matrix of elements 
\begin{eqnarray}\label{x_hoyw_def1}
({}_s\tilde{\mathbf{\Gamma}}^{Y}_v)_{i,j} = \gamma^Y(v-j, p+i-j)~\text{
while}~
_s\tilde{\mathbf{\gamma}}^{Y}_v = [\gamma^Y(v,p+1),\cdots,\gamma^Y(v,p+s)]'.
\end{eqnarray}
In the estimation, the ${\gamma}^Y(w,k)$ expression is replaced by its empirical version defined for a sample $y_1,\cdots, y_{NT}$ as
\begin{eqnarray}\label{emp_peacvf}
\hat{\gamma}^Y(w,k) = \frac{1}{N}\sum_{n=l}^r y_{nT+w} y_{nT+w-k},\end{eqnarray}
where\begin{eqnarray}
 l=\max\left(\left\lceil\frac{1-w}{T}\right\rceil,\left\lceil\frac{1-(w-k)}{T}\right\rceil\right), \quad  r=\min\left(\left\lfloor\frac{NT-w}{T}\right\rfloor,\left\lfloor\frac{NT-(w-k)}{T}\right\rfloor\right).
 \end{eqnarray}
The main idea of the presented method is finding an estimate of additive noise variance using high-order Yule-Walker equations and then utilizing it in low-order Yule-Walker equations for estimation of model's coefficients and innovations' variance. Let us define the functions $\mathbf{\Phi}^*_v(\cdot)$, $\sigma^{2*}_\xi(v)(\cdot)$ of the parameter $\sigma_Z^{2*}$ based on low-order Yule-Walker equations for $\{Y_t\}$:
\begin{eqnarray}\label{diversi1_phi}
\hat{\mathbf{\Phi}}^*_v(\sigma^{2*}_Z) = (\hat{\mathbf{\Gamma}}^Y_v-\sigma^{2*}_Z \mathbf{I}_p)^{-1}  \hat{\mathbf{\gamma}}^Y_v,~~
\hat{\sigma}^{2*}_\xi(v)(\sigma^{2*}_Z) = \hat{\gamma}^Y(v,0) - \hat{\mathbf{\Phi}}^*_v(\sigma^{2*}_Z)' \hat{\mathbf{\gamma}}^Y_v  - \sigma^{2*}_Z,
\end{eqnarray}
where $\hat{\mathbf{\Gamma}}^Y_v$ and $\hat{\mathbf{\gamma}}^Y_v$ are constructed using  Eq. \eqref{x_loyw_def1}  by replacing all {the} terms with their empirical counterparts,
\begin{eqnarray}\label{y_loyw_def1}
(\hat{\mathbf{\Gamma}}^Y_v)_{i,j} = \hat{\gamma}^Y(v-i,j-i),~~
\hat{\mathbf{\gamma}}^Y_v = [\hat{\gamma}^Y(v,1),\cdots,\hat{\gamma}^Y(v,p)]',
\end{eqnarray}
for $i,j=1,2,\cdots,p$. To construct an upper bound for $\sigma^{2*}_Z$, similarly as in \cite{diversi1}, let us also consider the following matrices $\hat{G}^Y_v$ for each $v=1,\cdots,T$:
\begin{eqnarray}\label{gY}
\hat{\mathbf{G}}_v^Y = 
\begin{bmatrix}
\hat{\gamma}^Y(v,0) & \hat{\mathbf{\gamma}}_v^{Y\prime} \\
\hat{\mathbf{\gamma}}^Y_v & \hat{\mathbf{\Gamma}}^Y_v
\end{bmatrix}.
\end{eqnarray}
As the estimated additive noise variance $\hat{\sigma}^{2*}_Z$ we consider the value which minimizes the following high-order Yule-Walker equations-based cost function
\begin{eqnarray}\label{jtotal}
J_{\text{total}}(\sigma^{2*}_Z) = \sum_{v=1}^T J_v(\sigma^{2*}_Z),
\end{eqnarray}
over an interval $\sigma^{2*}_Z \in [0,\zeta]$, for $\zeta = \min\{\min \text{eig}(\hat{\mathbf{G}}^Y_v), v = 1,\cdots,T\}$, where
\begin{eqnarray}\label{costhoyw}
 J_v({\sigma}^{2*}_Z) = || {}_s\hat{\tilde{\mathbf{\Gamma}}}^{Y}_v \hat{\mathbf{\Phi}}^*_v({\sigma}^{2*}_Z) - {}_s\hat{\tilde{\mathbf{\gamma}}}^{Y}_v ||^2_2,
\end{eqnarray}
\begin{eqnarray}\label{y_hoyw_def1}
({}_s\hat{\tilde{\mathbf{\Gamma}}}^{Y}_v)_{i,j} = \hat{\gamma}^Y(v-j,p+i-j),~~
_s\hat{\tilde{\mathbf{\gamma}}}^{Y}_v = [\hat{\gamma}^Y(v,p+1),\cdots,\hat{\gamma}^Y(v,p+s)]'
\end{eqnarray}
for $i=1,2,\cdots,s$, $j = 1,2,\cdots, p$. At the end, having the optimal $\hat{\sigma}^{2*}_Z$, for each $v=1,\cdots,T$ we calculate
\begin{eqnarray}
\hat{\mathbf{\Phi}}_v = (\hat{\mathbf{\Gamma}}^Y_v-\hat{\sigma}^{2}_Z \mathbf{I}_p)^{-1}  \hat{\mathbf{\gamma}}^Y_v,~
\hat{\sigma}^{2}_\xi(v) = \hat{\gamma}^Y(v,0) - \hat{\mathbf{\Phi}}_v' \hat{\mathbf{\gamma}}^Y_v - \hat{\sigma}^{2}_Z,
\end{eqnarray}
and estimate the variance of {the} innovations $\hat{\sigma}^{2}_\xi${, which is the mean of all the $\hat{\sigma}^{2}_\xi(v)$ across $v=1,\cdots,T$}. The entire procedure is presented in Algorithm \ref{algmethod3}.
\begin{algorithm}[H]
  \caption{Modified errors-in-variables method}\label{algmethod3}
  \begin{algorithmic}[1]
    \State Set value of $s$ (where $s\geq p$). In this paper, we set $s=p$.
    \State For each $v=1,\cdots,T$:
    \begin{enumerate}[itemsep=0pt,parsep=0pt,topsep=0pt,label=\footnotesize\roman*:]
    \item Construct $\hat{\mathbf{\Gamma}}^Y_v$ (Eq. \eqref{y_loyw_def1}), $\hat{\mathbf{\gamma}}^Y_v$ (Eq. \eqref{y_loyw_def1}), ${}_s\hat{\tilde{\mathbf{\Gamma}}}^{Y}_v$ (Eq. \eqref{y_hoyw_def1}) and ${}_s\hat{\tilde{\mathbf{\gamma}}}^{Y}_v$ (Eq. \eqref{y_hoyw_def1}).
    \item Construct $\hat{\mathbf{G}}^Y_v$ (Eq. \eqref{gY}) and compute $\min \text{eig}(\hat{\mathbf{G}}^Y_v)$.
    \end{enumerate}
    \State Compute $\zeta = \min \left\{ \min \text{eig}(\hat{\mathbf{G}}^Y_v),\: :\: v=1,\cdots,T \right\}$
    \State Determine $\hat{\sigma}^2_Z$ -- a value which minimizes $J_{\text{total}}(\sigma^{2*}_Z)$ (Eq. \eqref{jtotal}) over interval $\sigma^{2*}_Z \in [0, \zeta]$.
    \State For each $v=1,\cdots,T$ :
        \begin{enumerate}[itemsep=0pt,parsep=0pt,topsep=0pt,label=\footnotesize\roman*:]
        \item Compute $\hat{\mathbf{\Phi}}_v = (\hat{\mathbf{\Gamma}}^Y_v-\hat{\sigma}^{2}_Z \mathbf{I}_p)^{-1}  \hat{\mathbf{\gamma}}^Y_v$.
        \item Compute $\hat{\sigma}^{2}_\xi(v) = \hat{\gamma}^Y(v,0) - \hat{\mathbf{\Phi}}_v' \hat{\mathbf{\gamma}}^Y_v - \hat{\sigma}^{2}_Z$.
        \end{enumerate}
    \State Compute $\hat{\sigma}^2_{\xi} = 1/T\sum_{v=1}^T \hat{\sigma}^2_{\xi}(v)$
  \end{algorithmic}
\end{algorithm}

\end{document}